\documentclass[sigconf]{acmart}

\usepackage{amsmath,amssymb,amsfonts}
\usepackage{algorithmic}
\usepackage{graphicx}
\usepackage{textcomp}
\usepackage{xcolor}
\usepackage{booktabs}
\usepackage{colortbl}
\usepackage{multirow}
\usepackage{bm}
\usepackage[justification=centering]{caption}
\usepackage{balance}
\usepackage{float}
\usepackage{comment}
\usepackage{threeparttable}

\AtBeginDocument{%
  \providecommand\BibTeX{{%
    \normalfont B\kern-0.5em{\scshape i\kern-0.25em b}\kern-0.8em\TeX}}}

\copyrightyear{2021} 
\acmYear{2021} 
\setcopyright{acmcopyright}\acmConference[ESEM '21]{ACM / IEEE International Symposium on Empirical Software Engineering and Measurement (ESEM)}{October 11--15, 2021}{Bari, Italy}
\acmBooktitle{ACM / IEEE International Symposium on Empirical Software Engineering and Measurement (ESEM) (ESEM '21), October 11--15, 2021, Bari, Italy}
\acmPrice{15.00}
\acmDOI{10.1145/3475716.3475782}
\acmISBN{978-1-4503-8665-4/21/10}

\begin{document}

\title{Characteristics and Challenges of Low-Code Development: The Practitioners’ Perspective}




\author{Yajing Luo$^{1}$, Peng Liang$^{1*}$, Chong Wang$^{1}$, Mojtaba Shahin$^{2}$, Jing Zhan$^{3}$}
\affiliation{%
  \institution{$^{1}$School of Computer Science, Wuhan University, Wuhan, China}
  \institution{$^{2}$Faculty of Information Technology, Monash University, Melbourne, Australia}
  \institution{$^{3}$Department of Computer Science, University of Illinois at Urbana-Champaign, Champaign, United States}
  \institution{\{luoyajing, liangp, cwang\}@whu.edu.cn,  mojtaba.shahin@monash.edu,  jingz15@illinois.edu}
  \country{}
}

\renewcommand{\shortauthors}{Y. Luo et al.}

\begin{abstract}
\textbf{Background:} In recent years, Low-code development (LCD) is growing rapidly, and Gartner and Forrester have predicted that the use of LCD is very promising. Giant companies, such as Microsoft, Mendix, and Outsystems have also launched their LCD platforms. \textbf{Aim:} In this work, we explored two popular online developer communities, Stack Overflow (SO) and Reddit, to provide insights on the characteristics and challenges of LCD from a practitioners’ perspective. \textbf{Method:} We used two LCD related terms to search the relevant posts in SO and extracted 73 posts. Meanwhile, we explored three LCD related subreddits from Reddit and collected 228 posts. We extracted data from these posts and applied the Constant Comparison method to analyze the descriptions, benefits, and limitations and challenges of LCD. For platforms and programming languages used in LCD, implementation units in LCD, supporting technologies of LCD, types of applications developed by LCD, and domains that use LCD, we used descriptive statistics to analyze and present the results. \textbf{Results:} Our findings show that: (1) LCD may provide a graphical user interface for users to drag and drop with little or even no code; (2) the equipment of out-of-the-box units (e.g., APIs and components) in LCD platforms makes them easy to learn and use as well as speeds up the development; (3) LCD is particularly favored in the domains that have the need for automated processes and workflows; and (4) practitioners have conflicting views on the advantages and disadvantages of LCD. \textbf{Conclusions:} Our findings suggest that researchers should clearly define the terms when they refer to LCD, and developers should consider whether the characteristics of LCD are appropriate for their projects.
\end{abstract}

\ccsdesc[500]{Software and its engineering~Software development techniques}

\keywords{Low-Code Development, Stack Overflow, Reddit, Empirical Study}

\maketitle

\section{Introduction}
\label{sec:introduction}
With the growth of the Internet and the wave of digitalization, there is a growing need for enterprises to make quick and resilient responses to changing market requirements \cite{Sanchis2020Low}. According to
the research company Gartner, by 2021, the demand for
information systems will increase five times faster than the
ability to provide them by IT departments, because number of employees is not growing at a sufficient pace \cite{Waszkowski2019Low}. Furthermore, recruiting software engineers has become increasingly difficult as demand is high and supply is low \cite{torres2018demand}. In order to solve the problems above and adapt to this rapidly evolving world, companies are looking for quicker and cheaper ways to meet their software needs \cite{Fryling2019Low}. In response, low-code development (LCD) platforms have emerged with the promise that organizations can hire business professionals with no coding experience to build applications \cite{Fryling2019Low}.


The term ``low-code'' was first introduced to the public by Forrester Research in 2014 \cite{richardson2014new}, which states that firms prefer to choose low-code alternatives for fast, continuous, and test-and-learn delivery. The survey performed by Forrester \cite{richardson2016vendor} also shows that LCD platforms can accelerate development by 5 to 10 times. Moreover, these platforms also offer enterprises a more economical way to fulfil the market and/or enterprises internal requirements \cite{Sanchis2020Low}. Although LCD is booming in industry, there is no clear understanding of LCD as well as its practices. To this end, we plan to explore the characteristics and challenges of LCD from the perspective of practitioners.

To get practitioners' opinions, we have looked into online developer communities. As one of the online software development communities, Stack Overflow has been the most popular and widely used questions and answers (Q\&A) platform for developers to ask and answer questions since 2008 \cite{grant2013encouraging}. There are millions of software practitioners at SO who exchange knowledge, share experience, discuss problems they encountered during software development activities. Besides, Reddit is a network of subreddits based on people's interests, and members can submit content to the site, such as links, text posts, images, and videos, which are then voted up or down by other members. Furthermore, the posts are organized by subject into user-created boards called ``subreddits'', which cover a variety of topics including software development. In short, we collected data from both SO and Reddit, then we extracted the data items from the selected posts and analyzed them using the Constant Comparison method and descriptive statistic method.

\textbf{The contributions of this work}: (1) we explored how practitioners describe LCD based on their understanding; (2) we categorized and analyzed the features of LCD, including platforms, programming languages, implementation units, supporting technologies, application types, and domains of LCD; and (3) we also collected developers’ views on the strengths and weaknesses of LCD.

The paper is structured as follows: Related work is presented in Section \ref{sec:relatedWork}. Research questions and study design are then explained in Section \ref{sec:methodology}. The study results are provided in Section \ref{sec:results}, followed by a discussion of the results in Section \ref{sec:discussion}. The potential threats to the validity of the results are covered in Section \ref{sec:threats}, before concluding this work with future work directions in Section \ref{sec:conclusion}.
\section{Related Work}
\label{sec:relatedWork}

\subsection{Low-Code Development}
\label{sec:Low-Code Developmen}
Several studies have focused on the topic of LCD. Waszkowski \cite{Waszkowski2019Low} described the use of Aurea BPM low-code platform for automating business processes in manufacturing. Sahay \textit{et al.} \cite{Sahay2020Supporting} presented a technical survey of different LCD platforms based on a proposed conceptual comparative framework. Alamin \textit{et al.} \cite{Alamin2021An} also conducted an empirical study of developer discussions on the challenges of LCD, but the research questions between our study and Alamin \textit{et al.}'s work are totally different. Compared to Alamin \textit{et al.}'s research questions \cite{Alamin2021An} that focus on which LCD related topics are discussed in SO, how are these topics distributed across the LCD life cycle stages, and what LCD topics are the most difficult to answer, our study paid more attention to providing insights on the descriptions and characteristics of LCD, as well as the pros and cons brought by using LCD. Also, instead of obtaining data from one data source SO in Alamin \textit{et al.}'s work \cite{Alamin2021An}, we collected the data from both SO and Reddit to improve the external validity of the study results. Besides, Alamin \textit{et al.} \cite{Alamin2021An} used LCD platform names as tags (e.g., ``appmaker'' for Google Appmaker) to conduct a tag-based search in SO and collect the search results as their dataset. But we only conducted a keyword-based search in SO, as we thought that the posts collected with these LCD platform tags may be biased to specific LCD platforms which will threaten the generalizability of the results.

To the best of our knowledge, there are no studies that explore the understanding and description of LCD in practice. In our study, we came up with nine research questions that intend to provide a comprehensive overview of the characteristics of LCD, including its descriptions, platforms, programming languages, implementation units, supporting technologies, application types, domains, and benefits, as well as its challenges.

\subsection{Using Online Developer Communities in Software Engineering}
\label{sec:usingonlinecommunitiesinSE}
In recent years, millions of software developers are active in online developer communities to solve problems and share opinions, turning out to be great sources for researchers to study specific topics in software engineering. 

For example, Stack Overflow has become the most popular professional Q\&A website to gain answers and opinions from researchers and practitioners. Abdellatif \textit{et al.} \cite{Abdellatif2020Challenges} investigated the chatbot-related posts on Stack Overflow to pinpoint the major topics surrounding the discussions on the chatbot development. Chatterjee \textit{et al.} \cite{Chatterjee2020Finding} focused on Stack Overflow from the perspective of an individual seeking help with programming errors. Cummaudo \textit{et al.} \cite{Cummaudo2020Interpreting} used Stack Overflow to mine indications of the frustrations that developers appear to face when using computer vision services, while Zahedi \textit{et al.} \cite{Zahedi2020Mining} explored various aspects of continuous software engineering by mining related posts from Stack Overflow. In addition, other popular online developer communities are also used for researches in software engineering. For instance, Li \textit{et al.} \cite{Li2021Understanding} selected Reddit as one of the data sources for collecting discussions related to architecture erosion to look into the notion, causes, consequences, detection, and control of architecture erosion.

Given SO and Reddit contain a big volume of data based on the Q\&A mechanism that captures practitioners’ opinions, we decided to use these two online developer communities as our data sources to conduct the research on LCD.

\section{Methodology}
\label{sec:methodology}

\begin{figure*}[h]
	\centering
	\includegraphics[width=\linewidth]{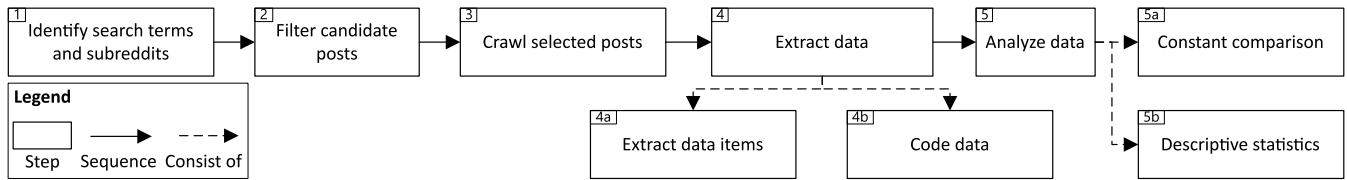}
	\caption{Overview of the research process}
	\label{fig:Overview of research process}
\end{figure*}

We set the goal of this study based on the Goal-Question-Metric approach \cite{Basili1994TheGQ}: \textbf{analyze} the perception of LCD in industry \textbf{for the purpose of} characterizing \textbf{with respect to} various aspects of LCD \textbf{from the point of view of} practitioners \textbf{in the context of} posts concerned with LCD from SO and Reddit. In the following subsections, we explain the Research Questions (RQs), their rationale, and the research process (see Figure \ref{fig:Overview of research process}) used to answer the RQs.\\

\subsection{Research Questions} \label{Research Questions}

\noindent \textbf{RQ1: What is the understanding of LCD?}\\
\noindent \emph{Rationale:} Practitioners have different understandings about LCD according to their experience and expectations, and also use different synonyms to refer to LCD in practices, such as ``no-code development'' and ``zero-code development''. This RQ aims to get an understanding of LCD from practitioners.\\
\noindent \textbf{RQ2: What platforms are used in LCD?}\\
\noindent \emph{Rationale:} LCD platforms provide an integrated development environment dedicated to LCD. There exist many platforms which offer various features to LCD. This RQ intends to identify popular LCD platforms along with their features.\\
\noindent \textbf{RQ3: What programming languages are used in LCD?}\\
\noindent \emph{Rationale:} LCD helps practitioners write less code while the developers sometimes still need to do some hand-coding. This RQ explores the programming languages that have been used for various purposes (e.g., customizing features) in LCD.\\
\noindent \textbf{RQ4: What are the major implementation units in LCD?}\\
\noindent \emph{Rationale:} This RQ intends to understand how low the code is in LCD. In other words, what high level of implementation units (e.g., component-based unit and framework-based unit) the development is based on.\\
\noindent \textbf{RQ5: What are the supporting technologies used in LCD?}\\
\noindent \emph{Rationale:} To provide convenience with minimum hand-coding and other advantages, LCD needs to employ certain technologies running underneath. This RQ aims to identify the supporting technologies used in LCD.\\
\noindent \textbf{RQ6: What types of applications are developed by LCD?}\\
\noindent \emph{Rationale:} Many LCD platforms have been available on the market and used in industry. These platforms might be appropriate for developing different types of applications (e.g., mobile applications, desktop applications, and web applications). This RQ intends to get an overview of the application types that are developed by LCD.\\
\noindent \textbf{RQ7: What are the domains that use LCD?}\\
\noindent \emph{Rationale:} LCD has been used to develop applications from various domains (e.g., finance, telecommunication). This RQ intends to investigate the domains of these applications, which are appropriate to be developed using LCD.\\
\noindent \textbf{RQ8: What are the benefits of LCD?}\\
\noindent \emph{Rationale:} LCD as an emerging software development paradigm can lead to many benefits (e.g., decreased time and cost). We ask this question with the purpose of collecting the benefits brought to the system development by applying LCD.\\
\noindent \textbf{RQ9: What are the limitations and challenges of LCD?}\\
\noindent \emph{Rationale:} LCD is still in its early stage of practices with many limitations, and it also raises problems that are not encountered during traditional development. The answer to this RQ will help practitioners make informed decisions when deciding whether or not to employ LCD in their projects.\\

\subsection{Data Collection and Filtering}
\label{sec:subject_projects} 
To answer the RQs, we need to collect the knowledge, opinions, and experience from practitioners about LCD. As discussed in Section \ref{sec:usingonlinecommunitiesinSE}, SO and Reddit contain a large amount of Q\&A-based data that captures the above aspects from practitioners, we decided to choose the two websites as the data sources for our study and conducted the data collection from January 2021 to April 2021.\\

\subsubsection{Search and collect candidate posts}
We started by searching the existing literature about ``low-code'' on the Internet. We identified the synonyms (i.e., ``low-code'', ``zero-code'', and ``no-code'') as search terms.

\textbf{For SO}, during the pilot search process, we found that some practitioners use LCD platform names as tags in SO, e.g., ``outsystems'' for the OutSystems platform\footnote{https://www.outsystems.com/}. However, we thought that the introduction of LCD platform names as search terms is biased to specific LCD platforms and thus threatens the search results. Therefore, we decided not to include these LCD platform names in the search terms.

During the keyword-based search process at SO, as the effectiveness of a keyword-based mining approach largely depends on having a proper set of keywords, we applied the systematic approach conducted by Bosu \textit{et al.} \cite{Bosu2014Identifying} to identity the keywords required in our study. The steps are the following (the NLTK package\footnote{http://www.nltk.org/} was used in several steps below):
\begin{enumerate}
\item We created an initial set of keywords (i.e., ``low-code'', ``zero-code'', and ``no-code'').
\item We searched the SO community using the initial set of keywords to build a corpus of documents, including posts that contain at least one keyword.
\item We converted all words to lowercase. Then we cleaned the corpus by removing all the punctuation and digits. Next, we deleted stopwords from the corpus using the NLTK library. Finally, we created a list of tokens. 
\item We applied the Porter stemming algorithm \cite{Willett2006The} to obtain the stem of each token (e.g., \textit{plays}, \textit{played}, \textit{playing} all became \textit{play}).
\item We created a Document-Term matrix \cite{Tan2005Introduction} from the corpus.
\item We determined whether any additional words that co-occur frequently with each of our initial keywords. If the co-occurrence probability is higher than 0.05, it can be considered for inclusion in the keywords set.
\item We manually analyzed the list of frequently co-occurring additional words to identify whether to include any of them in the final keywords set.
\end{enumerate}

After performing the steps above, we did not find any additional keywords that co-occur with each of our initial keywords. So we considered our original keyword set to be adequate and comprehensive.

Next, to make the results obtained from the search more precise, we refined and narrowed the search terms. As SO provides the tip that we can enter the keywords in quotes to search a specific phrase, we initially quoted the search terms during the search. After a cursory reading of the search results, we found that some irrelevant results were included. Hence, we excluded the irrelevant results by prefixing the according terms with ``-'' in our search query. For example, we discovered a number of ``low code coverage'' related posts in the ``low code'' search results that were not relevant to our topic. So, we changed the search string to ``low code''-``low code coverage'', which excluded ``low code coverage'' related posts. 

Then we conducted a pilot data extraction. The first author randomly selected 10 posts from the search results of all search terms. The purpose was to verify whether our selected search terms were appropriate and whether the data items could be extracted from the posts obtained from the search. Next, the first author extracted data from these 10 posts and then discussed the extraction results with the second author to see if the extraction results were correct. Moreover, since we got 19,455 retrieved results by the search term ``no-code'', and we checked the first 500 retrieved posts and found only 6 LCD relevant posts (1.2\%), considering that SO ranks the search results according to their relevance to the search term, we decided to remove the posts obtained by the search term ``no-code'' in the data collection step due to the low return on investment. The search terms used and their numbers of retrieved posts from SO are listed in Table \ref{Search terms on Stack Overflow}.

\textbf{For Reddit}, after exploring the subreddits on Reddit, we finally found three subreddits named ``Low Code''\footnote{https://www.reddit.com/r/lowcode/}, ``No Code''\footnote{https://www.reddit.com/r/nocode/}, and ``nocodelowcode''\footnote{https://www.reddit.com/r/nocodelowcode/}, which contain questions and discussions about LCD. We then chose all the posts in these three subreddits as part of our data sources. The subreddits and their numbers of posts from Reddit are listed in Table \ref{Subreddits in Reddit}.

\subsubsection{Filter candidates}
The criteria for filtering posts are defined as follows: (I1) If a post contains at least one data item to be extracted as listed in Table \ref{Data Items and RQ}, we include it. (E1) If the topic of a post is related to LCD but does not draw any useful information, we exclude it. (E2) If a post contains only videos or links and no data items that can be extracted, we exclude it (we do not extract data from the videos or links). This step was conducted by the first author since LCD is a simple concept and these criteria are easy to check, and only a few posts that the first author could not decide were discussed with the second author to reach a decision. After excluding the irrelevant posts in the search results, we finally got 301 LCD related posts out of 1574 posts from SO and Reddit. The details are shown in Table \ref{Search terms on Stack Overflow} and Table \ref{Subreddits in Reddit} for SO and Reddit, respectively.

\begin{table}[h]
\caption{Search terms used in Stack Overflow}
\label{Search terms on Stack Overflow}
\resizebox{\columnwidth}{!}{%
\begin{threeparttable}
\begin{tabular}{@{}cccc@{}}
\toprule
\textbf{\#} & \textbf{Search Term} & \textbf{\begin{tabular}[c]{@{}c@{}}Number of \\ Retrieved Posts\end{tabular}} & \textbf{\begin{tabular}[c]{@{}c@{}}Number of \\ Selected Posts\end{tabular}} \\ \midrule
\textbf{ST1}          & ``low code''-``low code coverage''                  & 127              & 46\\
\textbf{ST2}          & ``zero code''-``non zero code''-                    & 402              & 27\\
\textbf{}           & ``zero code coverage''-``zero code change''-        &                  &\\
\textbf{}           & ``zero code changes''-``zero code solution''        &                  &\\ \hline
\textbf{Total}      &                                                 & 529              & 73\\ \bottomrule
\end{tabular}
\begin{tablenotes}
        \footnotesize
        \item[1] Note that the hyphen ``-'' in the search terms has no effect on the search results at SO. For example, the search terms ``low-code'' and ``low code'' get the same search results.
      \end{tablenotes}
  \end{threeparttable}
}
\end{table}

\begin{table}[h]
\caption{Subreddits used in Reddit}
\label{Subreddits in Reddit}
\resizebox{\columnwidth}{!}{%
\begin{tabular}{@{}cccc@{}}
\toprule
\textbf{\#}    & \textbf{Subreddit} & \textbf{Number of Posts} & \textbf{Number of Selected Posts} \\ \midrule
\textbf{SR1}     & Low Code            & 52                                 & 23                                \\
\textbf{SR2}     & No Code             & 937                                & 199                               \\
\textbf{SR3}     & nocodelowcode       & 56                                 & 6                                 \\ \midrule
\textbf{Total} &                     & 1045                               & 228                               \\ \bottomrule
\end{tabular}
}
\end{table}

\subsection{Data Extraction and Analysis} 

\subsubsection{Extract data}
To answer the RQs in Section \ref{Research Questions}, we extracted the data items as listed in Table \ref{Data Items and RQ} from the selected posts. During this process, the first and fifth authors conducted a pilot data extraction with 10 posts independently, and any inconsistent extraction results were discussed with the second author to get a consensus. They further extracted data according to the data items from the rest selected posts, marked the uncertain paragraphs, and then discussed them with the second author. The first author re-examined the extraction results of all the posts to make sure that all the data were extracted correctly.

\begin{table}[h]
\caption{Data items extracted and their corresponding RQs}
\label{Data Items and RQ}
\resizebox{\columnwidth}{!}{
\begin{tabular}{|c|l|l|c|}
\hline
\textbf{\#} & \textbf{Data item}                                                           & \textbf{Description}                                                     & \textbf{RQ} \\ \hline
D1          & Description of LCD                                                           & \textit{\begin{tabular}[c]{@{}l@{}}Descriptions of LCD by practitioners \\ based on their understanding\end{tabular}} & RQ1         \\ \hline
D2          & Platforms used in LCD                                                        & \textit{The platforms that are used in LCD}                                                                           & RQ2         \\ \hline
D3          & \begin{tabular}[c]{@{}l@{}}Programming languages \\ used in LCD\end{tabular} & \textit{\begin{tabular}[c]{@{}l@{}}Programming languages that are \\ used in LCD\end{tabular}}                        & RQ3         \\ \hline
D4          & \begin{tabular}[c]{@{}l@{}}Implementation Unit in \\ LCD\end{tabular}        & \textit{\begin{tabular}[c]{@{}l@{}}The implementation units on which \\ LCD is based\end{tabular}}                    & RQ4         \\ \hline
D5          & \begin{tabular}[c]{@{}l@{}}Supporting technologies \\ of LCD\end{tabular}    & \textit{\begin{tabular}[c]{@{}l@{}}Supporting technologies behind \\ the LCD\end{tabular}}                            & RQ5         \\ \hline
D6          & Application Type                                                             & \textit{\begin{tabular}[c]{@{}l@{}}Types of applications developed \\ by LCD\end{tabular}}                            & RQ6         \\ \hline
D7          & Domain                                                                       & \textit{Domains that use LCD}                                                                                         & RQ7         \\ \hline
D8          & Benefits of LCD                                                              & \textit{Benefits brought by LCD}                                                                                      & RQ8         \\ \hline
D9          & \begin{tabular}[c]{@{}l@{}}Limitations and \\ challenges of LCD\end{tabular} & \textit{\begin{tabular}[c]{@{}l@{}}Existing shortcomings and future \\ challenges of LCD\end{tabular}}                & RQ9         \\ \hline
\end{tabular}
}
\end{table}

\subsubsection{Analyze data}
For RQ1, RQ8, and RQ9, we applied the Constant Comparison method \cite{Glaser1965The} for qualitative data analysis. Our analysis process involved the following steps: (1) The first author collected data from the selected posts (including titles, questions, comments, and all the answers) through data crawling. (2) The first author labeled these posts with codes that succinctly summarize the data items (for answering RQs). (3) The first author grouped all the codes into higher-level concepts and turned them into categories, and then the second author checked the coding results, and any divergence in the coding and categorization results were further discussed until the two authors reached an agreement. To effectively code and categorize data, we used the qualitative data analysis tool MAXQDA\footnote{https://www.maxqda.com/}. For the other RQs, we used descriptive statistics to analyze and present the results. The data analysis methods used for the data items and their corresponding RQs are listed in Table \ref{Data Items and the data analysis methods used for the research questions}. All the data and coding results of this study have been provided online for replication purpose \cite{replpack}.

\begin{table}[h]
\caption{Data items and their analysis methods for answering the RQs}
\label{Data Items and the data analysis methods used for the research questions}
\resizebox{\columnwidth}{!}{
\begin{tabular}{|c|l|l|c|}
\hline
\textbf{\#} & \textbf{Data item}                                                           & \textbf{Data analysis method} & RQ  \\ \hline
D1          & Description of LCD                                                           & Constant comparison                                & RQ1 \\ \hline
D2          & Platform used in LCD                                                        & Descriptive statistics                             & RQ2 \\ \hline
D3          & \begin{tabular}[c]{@{}l@{}}Programming language used in \\ LCD\end{tabular} & Descriptive statistics                             & RQ3 \\ \hline
D4          & Implementation unit in LCD                                                   & Descriptive statistics                             & RQ4 \\ \hline
D5          & Supporting technology of LCD                                               & Descriptive statistics        & RQ5 \\ \hline
D6          & Application Type                                                             & Descriptive statistics                             & RQ6 \\ \hline
D7          & Domain                                                                       & Descriptive statistics                             & RQ7 \\ \hline
D8          & Benefit of LCD                                                     & Constant comparison                                & RQ8 \\ \hline
D9          & \begin{tabular}[c]{@{}l@{}}Limitation and challenge of \\ LCD\end{tabular} & Constant comparison                                & RQ9 \\ \hline
\end{tabular}
}
\end{table}


\section{Results}
\label{sec:results}
In this section, we present the results of the nine RQs. For each RQ, we first explain the analysis approach that was employed to answer the question, then we present the results.\\

\noindent \textbf{RQ1: What is the understanding of LCD?}\\
To answer RQ1, we applied the Constant Comparison method to analyze the extracted data. We finally extracted 106 instances that practitioners use to describe LCD according to their understanding and categorized them into 11 types as listed in Table \ref{Terms that practitioners used to describe LCD}.

We found that most practitioners tend to use \textit{low-code} (e.g., ``\textit{The coding effort is low}'') to describe LCD. In other words, they think that the coding effort is low in LCD. The term \textit{drag and drop} comes second, followed closely by \textit{visual programming}. Some practitioners also use \textit{pre-designed templates}, \textit{non-professional programmers friendly}, \textit{what you see is what you get (WYSIWYG)}, and \textit{business process} to demonstrate their understanding and perception of LCD. A few others consider that LCD utilizes a \textit{graphical user interface} to develop programs, and one use case is \textit{build automation} to ``\textit{automate unattended operations with minimal human involvement}
''. They also think that LCD brings convenience to \textit{database operations}. For example, one practitioner mentioned that ``\textit{It provides some cool tools for generating CRUD entities by scaffolding}''. Only one developer commented in the post that LCD ``\textit{combines visual and code workflows to facilitate collaboration in the same environment}''.

\begin{table}[h]
\caption{Terms that practitioners use to describe LCD}
\label{Terms that practitioners used to describe LCD}
\resizebox{\columnwidth}{!}{
\begin{tabular}{|l|l|c|}
\hline
\textbf{Term}                                                                    & \textbf{Example}                                                                                                                                                                                                                                      & \textbf{Count} \\ \hline
Low-code                                                                         & \textit{\begin{tabular}[c]{@{}l@{}}You need less programming skills and \\ you are able to realize your processes \\ without the need of coding\end{tabular}}                                                                                                             & 24             \\ \hline
Drag and drop                                                                    & \textit{\begin{tabular}[c]{@{}l@{}}can probably do everything through \\ drag-and-drop\end{tabular}}                                                                                                                                                                       & 23             \\ \hline
Visual programming                                                               & \textit{\begin{tabular}[c]{@{}l@{}}low-code is a visual approach to \\ software development\end{tabular}}                                                                                                                                                                  & 16             \\ \hline
\begin{tabular}[c]{@{}l@{}}Pre-designed \\ templates\end{tabular}                & \textit{\begin{tabular}[c]{@{}l@{}}gives everyone from business users to \\ advanced developers the right automation \\ canvas to build great software robots\end{tabular}}                                                                                                & 10             \\ \hline
\begin{tabular}[c]{@{}l@{}}Non-professional \\ programmers friendly\end{tabular} & \textit{\begin{tabular}[c]{@{}l@{}}This is especially useful for people with \\ limited coding skills or devs that want to \\ automate something quickly while not having \\ to think about all aspects of development, \\ such as deployment, security, ...\end{tabular}} & 9              \\ \hline
\begin{tabular}[c]{@{}l@{}}What you see is \\ what you get\end{tabular}          & \textit{meant for WYSIWYG app maker}                                                                                                                                                                                                                                       & 7              \\ \hline
Business process                                                                 & \textit{especially designed for process owners}                                                                                                                                                                                                                            & 6              \\ \hline
\begin{tabular}[c]{@{}l@{}}Graphical user \\ interface\end{tabular}              & \textit{They provide you with a graphical wizard}                                                                                                                                                                                                                        & 5              \\ \hline
Build automation                                                                 & \textit{\begin{tabular}[c]{@{}l@{}}automate unattended operations with minimal \\ human involvement\end{tabular}}                                                                                                                                                          & 3              \\ \hline
Database operation                                                               & \textit{\begin{tabular}[c]{@{}l@{}}It provides some cool tools for generating \\ CRUD entities by scaffolding\end{tabular}}                                                                                                                                               & 2              \\ \hline
\begin{tabular}[c]{@{}l@{}}Collaboration in the \\ same environment\end{tabular} & \textit{\begin{tabular}[c]{@{}l@{}}combines visual and code workflows allowing \\ designers, developers, and low-code users \\ to work together in a single environment\end{tabular}}                                                                                     & 1              \\ \hline
\end{tabular}
}
\end{table}

\vspace{0.2cm}

\noindent \textbf{RQ2: What platforms are used in LCD?}\\
To answer RQ2, we manually identified platforms used in LCD from all the posts, then we analyzed them. During the identification process, if one platform was mentioned multiple times in the same post, it is only counted once. We counted the number of times that different platforms were mentioned in the posts, listed the companies that develop them, and explored whether they are open-source or commercial. The platforms that appear 10 times or more are presented in Table \ref{Platforms used in LCD, their companies, whether open-source or not and counts}. We also classified the platforms according to the parts of applications developed by LCD as shown in Table \ref{Platforms classified by the parts of applications developed by LCD}. 

Regarding Table \ref{Platforms used in LCD, their companies, whether open-source or not and counts}, of the 21 LCD platforms we listed (mentioned more than 10 times), 14 of them are commercial while 7 are open-source. Regarding Table \ref{Platforms classified by the parts of applications developed by LCD}, we identified 137 platforms (mentioned at least once), which are categorized into seven parts of applications developed by LCD with \textit{Not mentioned} and \textit{Others}. \textit{Not mentioned} means that we did not find the answers from those platforms' official websites or developers' descriptions. \textit{Others} means that those platforms do not belong to any of the above classifications. Note that one platform can be used to develop multiple parts with LCD and therefore can be classified into multiple categories (e.g., Airtable, Bubble.io, and Webflow), and the total number of counts (218) is more than the number of platforms (137).

\begin{table}[h]
\caption{Platforms used in LCD, their companies, open-source or commercial, and counts}
\label{Platforms used in LCD, their companies, whether open-source or not and counts}
\resizebox{\columnwidth}{!}{
\begin{tabular}{|l|l|l|c|}
\hline
\textbf{Platform name} & \textbf{Company name}               & \textbf{Open-source or commercial} & \textbf{Count} \\ \hline
Bubble.io              & \textit{Bubble}                     & Commercial                         & 96             \\ \hline
Webflow                & \textit{Webflow Inc.}               & Commercial                         & 63             \\ \hline
Adalo                  & \textit{Adalo}                      & Commercial                         & 50             \\ \hline
Airtable               & \textit{Airtable}                   & Commercial                         & 45             \\ \hline
Appgyver               & \textit{Appgyver}                   & Open-source                        & 38             \\ \hline
Glide                  & \textit{Glide}                      & Open-source                        & 27             \\ \hline
Wix (Editor X)         & \textit{Wix.com Inc.}               & Commercial                         & 19             \\ \hline
Power Apps             & \textit{Microsoft}                  & Commercial                         & 18             \\ \hline
Zapier                 & \textit{Zapier}                     & Commercial                         & 17             \\ \hline
DronaHQ                & \textit{Deltecs InfoTech Pvt. Ltd.} & Commercial                         & 17             \\ \hline
WordPress              & \textit{WordPress Foundation}       & Open-source                        & 16             \\ \hline
Softr.io               & \textit{Softr}                      & Commercial                         & 16             \\ \hline
Backendless            & \textit{Backendless}                & Open-source                        & 16             \\ \hline
Appsheet               & \textit{Google}                     & Commercial                         & 16             \\ \hline
Outsystems             & \textit{Outsystems}                 & Commercial                         & 12             \\ \hline
Thunkable              & \textit{Thunkable}                  & Commercial                         & 11             \\ \hline
Draftbit               & \textit{Draftbit}                   & Open-source                        & 11             \\ \hline
Xano                   & \textit{Xano Inc.}                  & Commercial                         & 10             \\ \hline
Wappler.io             & \textit{Wappler}                    & Open-source                        & 10             \\ \hline
Shopify                & \textit{Shopify}                    & Commercial                         & 10             \\ \hline
Integromat             & \textit{Celonis GmbH}               & Open-source                        & 10             \\ \hline
\end{tabular}
}
\end{table}

\begin{table}[]
\caption{Platforms classified by the parts of applications developed by LCD}
\label{Platforms classified by the parts of applications developed by LCD}
\resizebox{\columnwidth}{!}{
\begin{tabular}{|l|l|c|}
\hline
\textbf{\begin{tabular}[c]{@{}l@{}}The part of \\ applications \\ developed by LCD\end{tabular}} & \textbf{Platform name}                                                                                                                                                                                                                                                                                                                                                                                                                                                                                                                                                                                                                                                     & \textbf{Count} \\ \hline
Frontend                                                                                         & \textit{\begin{tabular}[c]{@{}l@{}}Adalo, Alpha Software, Andromo, Anvil, App maker, \\ Appgyver, Appian, Apprat, Appsheet, AppyPie, \\ Backendless, BettyBlocks, Bildr, Bravo Studio, Bubble.io, \\ CalcuBuilder, Carrd, Caspio, Draftbit, DrapCode, Dribble, \\ DronaHQ, Elementor, Expression Blend, Figma, Fliplet, \\ FlutterFlow, Kony Visualizer, Mendix, Noloco, \\ Outsystems, PixelCraft, Pory.io, Power Apps, pxCode, \\ Reach.at, Retool, Softr.io, Stacker, Storyboard, \\ Tadabase, Thunkable, UI Bakery, Undaku, Unqork, V One, \\ Verastream Host Integrator, Wappler.io, WaveMaker, \\ Webase, Webflow, Weflow, WeWeb, Wix, WordPress, Zyro\end{tabular}} & 56                                  \\ \hline
Workflow                                                                                         & \textit{\begin{tabular}[c]{@{}l@{}}Activiti, Airtable, Amazon Honeycode, Appian, \\ Automate.io, AwareIM, Azure Logic Apps, Bonitasoft, \\ Boomi Flow, Bubble.io, Budibase, Camunda, DronaHQ, \\ Drupal, Google Tables,Parabola, Intalio, Integromat, \\ jBPM, Joget, Knack, MakerPad, Mendix, N8n, Node-RED, \\ Outsystems, Pega, Pory.io, Power Apps, Power Automate, \\ Quickbase, Salesforce, ServiceNow, Slingr, Stackby, \\ Tadabase, TrackVia, Twilio Studio, UiPath Apps, \\ Undaku, UnifiedAI, Unqork, Webflow, Zapier, Zoho\end{tabular}}                                                                                                                        & 44                                  \\ \hline
Integration                                                                                      & \textit{\begin{tabular}[c]{@{}l@{}}Adalo, Airtable, Alpha Software, Appgyver, Appian, \\ AppMaker, Appsheet, AppyPie, AwareIM, Backendless, \\ BettyBlocks, Bildr, Bravo Studio, Bubble.io, Caspio, \\ Draftbit, DronaHQ, FlutterFlow, Knack, Memberstack, \\ Mendix, NocodeAPI, Parabola, Pory.io, Power Apps, \\ PrestoAPI, Quickbase, Retool, Slingr, Stacker, \\ Syndesis, Tadabase, TrackVia, UI Bakery, Unqork, \\ V One, WaveMaker, WeWeb, WordPress, Zyro\end{tabular}}                                                                                                                                                                                            & 40                                  \\ \hline
Backend                                                                                          & \textit{\begin{tabular}[c]{@{}l@{}}8base, Adalo, Alpha Software, Andromo, App maker, \\ Appgyver, Apprat, Appsheet, AppyPie, Backendless, \\ Bravo Studio, Bubble.io, Byteline, Caspio, Draftbit, \\ DrapCode, DronaHQ, Easy Tables, Firebase, flutterflow, \\ Kelp, Kinvey, Linx, Mendix, Noloco, Outsystems, \\ Power Apps, PrestoAPI, Thunkable, WaveMaker, Webase, \\ Webflow, WordPress, Xano, Zyro\end{tabular}}                                                                                                                                                                                                                                                     & 35                                  \\ \hline
Framework                                                                                        & \textit{\begin{tabular}[c]{@{}l@{}}Appsmith, Bildr, funkLang, Glide, Lowdefy, Orientation \\ Aware Control, Picocli, QTKit, Remake, Sonata Admin, \\ Substack, Wappler.io, WordPress\end{tabular}}                                                                                                                                                                                                                                                                                                                                                                                                                                                                         & 13             \\ \hline
Database operations                                                                              & \textit{\begin{tabular}[c]{@{}l@{}}8base, Alfresco, AwareIM, JayStack, Jhipster, Loopback, \\ Sonata Admin, Stackby, Tadabase , TeamDesk, Webase, \\ Zenbase\end{tabular}}                                                                                                                                                                                                                                                                                                                                                                                                                                                                                                 & 12                                  \\ \hline
Data visualization                                                                               & \textit{Google Data Studio, Kelp, Qlik, Tableau}                                                                                                                                                                                                                                                                                                                                                                                                                                                                                                                                                                                                                           & 4                                   \\ \hline
Not mentioned                                                                                    & \textit{\begin{tabular}[c]{@{}l@{}}Brizy.cloud, Internal.io, Open Lowcode, Sharetribe, \\ Skyve, Squarespace\end{tabular}}                                                                                                                                                                                                                                                                                                                                                                                                                                                                                                                                                 & 6                                   \\ \hline
Others                                                                                           & \textit{\begin{tabular}[c]{@{}l@{}}Boundless (Labs), Dialogflow, Godot engine, REI3, \\ Shopify, Unity, Unreal Engine, WooCommerce\end{tabular}}                                                                                                                                                                                                                                                                                                                                                                                                                                                                                                                           & 8                                   \\ \hline
\end{tabular}
}
\end{table}

\vspace{0.2cm}

\begin{figure}[h]
	\centering
	\includegraphics[width=1.0\linewidth]{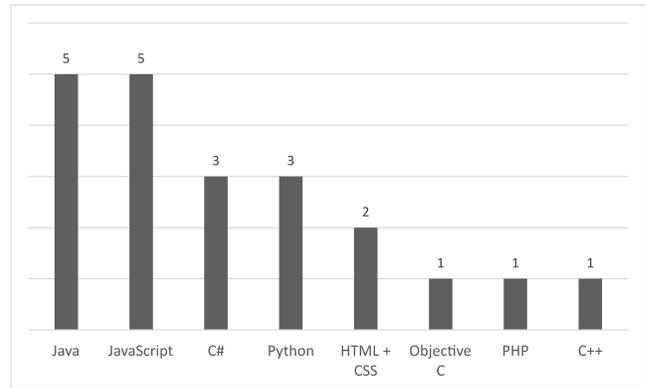}
	\caption{Programming languages used in LCD}
	\label{fig:Programming language used in LCD}
\end{figure}

\noindent \textbf{RQ3: What programming languages are used in LCD?}\\
To answer RQ3, we extracted the data ``programming languages used in LCD'' from the related posts and counted them by descriptive statistics. Figure \ref{fig:Programming language used in LCD} shows the statistical result. 

In total, we found 21 posts that mention the programming languages developers use for LCD. Five developers mentioned that they practiced LCD with \textit{Java}, as one post said ``\textit{I'm working with an enterprise `low-code'-tool to build applications for our company. Some parts can be written in Java and other languages}'', while another five practitioners said that they used \textit{Javascript} for LCD. There are also three developers used \textit{C\#} and \textit{Python} in LCD, respectively. Two developers commented that they used LCD with the combination of \textit{HTML and CSS}, for example, one developer stated that ``\textit{My recommendation would be to use webflow - it's simple to use for HTML and CSS}''. \textit{Objective C}, \textit{PHP}, and \textit{C++} are each mentioned in one post as being used for LCD.

\vspace{0.2cm}

\begin{figure}[h]
	\centering
	\includegraphics[width=1.0\linewidth]{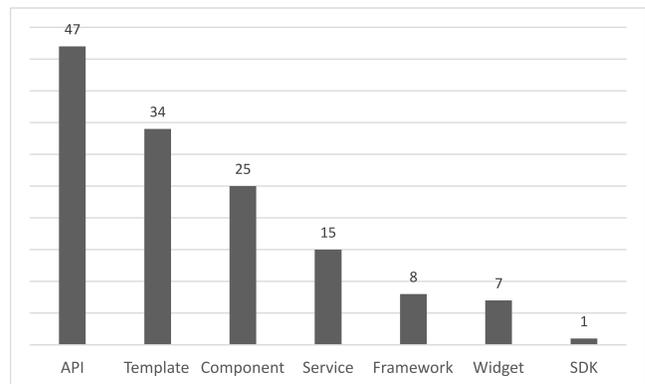}
	\caption{Implementation units in LCD}
	\label{fig:Implementation Unit in LCD}
\end{figure}

\noindent \textbf{RQ4: What are the major implementation units in LCD?}\\
To answer RQ4, we explored the implementation units that the LCD platforms collected in RQ2 are based on and counted these units by descriptive statistics. Figure \ref{fig:Implementation Unit in LCD} presents the statistical result. 

We extracted 137 instances from all the posts, 47 of which are \textit{API-based}, as one post stated ``\textit{What the app does is to call different Salesforce (low-code platform) APIs, both REST and SOAP, based on the input parameters AND based on the responses of some of the previous calls}'', while 34 are \textit{template-based}. There are 25 LCD platforms that are \textit{component-based}. Among the remaining 31 platforms mentioned, 15 LCD platforms are based on \textit{services}, 8 are based on \textit{frameworks}, 7 are based on \textit{widgets}, and the rest one is based on \textit{SDK}. As an example of framework-based LCD platforms, one post said that ``\textit{Loopback is an awesome framework that offers a full REST API to all CRUD available operations with zero code.}''

\vspace{0.2cm}

\noindent \textbf{RQ5: What are the supporting technologies used in LCD?}\\
To answer RQ5, we finally got 7 posts out of 301 posts containing supporting techniques behind LCD.

Two posts mention that LCD is based on ``\textit{React Native}'' as one post said that ``\textit{Draftbit exports code using Expo which is an SDK based on React Native}'', and the other post commented that ``\textit{open source nocode platform with drag and drop interface, built on top of nodejs}''. A post specified that LCD platform ``\textit{runs spark underneath}'', while another posted highlighted that LCD ``\textit{contains some advanced server-side scripting techniques}''. There is also a post stating that the LCD platform ``\textit{offers a SOAP connector for consuming SOAP web services}'', while another post claimed that ``\textit{they use JSONata to translate data schemas between platforms}''.

\vspace{0.2cm}

\noindent \textbf{RQ6: What types of applications are developed by LCD?}\\
To answer RQ6, we identified the application types that appear in the posts to get an overview of types of applications developed by LCD. we ended up with 70 posts mentioning the types of applications that they were developing or had developed.

\begin{figure}[h]
	\centering
	\includegraphics[width=1.0\linewidth]{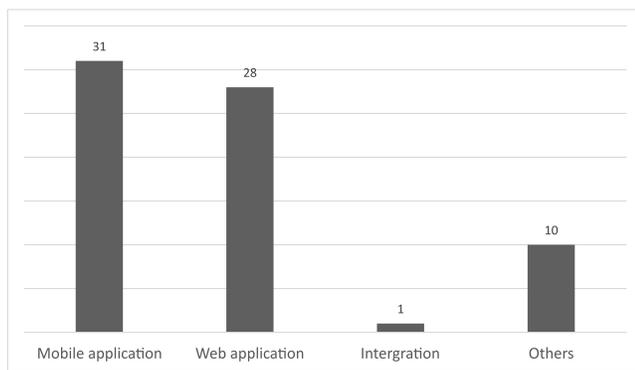}
	\caption{Application types developed by LCD}
	\label{fig:Application type}
\end{figure}

As shown in Figure \ref{fig:Application type}, 31 \textit{mobile applications}, 28 \textit{web applications}, and 1 \textit{integration application} were developed through LCD. There are 10 applications that are classified as \textit{Others}, because the posts only mentioned ``application'' without further information, and they cannot be explicitly classified as one of the above types.

\vspace{0.2cm}

\begin{figure}[h]
	\centering
	\includegraphics[width=1.0\linewidth]{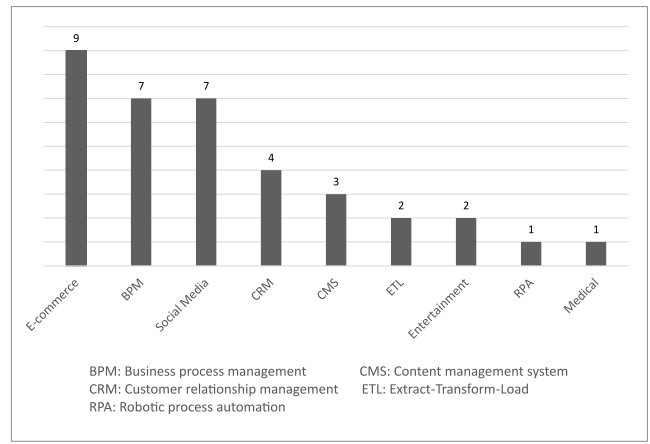}
	\caption{Application domains that use LCD}
	\label{fig:Domain}
\end{figure}

\noindent \textbf{RQ7: What are the domains that use LCD?}\\
To answer RQ7, we counted the domains that use LCD and got the result as shown in Figure \ref{fig:Domain}. 36 instances covering nine domains have been extracted.

The result shows that the domain where LCD is most used is \textit{E-commerce} including business-to-business (B2B) and business-to-consumer (B2C), and it was mentioned 9 times. \textit{Business Process Management (BPM)} and \textit{Social Media} were mentioned 7 times each, with \textit{Social Media} specifically containing partying, chatting, dating, and blogging applications, etc. \textit{Customer Relationship Management (CRM)} was talked about 4 times, followed by \textit{Content Management System (CMS)} being discussed 3 times. \textit{Extract-Transform-Load (ETL)} and \textit{Entertainment} were presented 2 times, respectively. In the entertainment domain, there are posts concerning the use of low-code game engines to develop games. At last, \textit{Robotic Process Automation (RPA)} and \textit{Medical} were mentioned once each.

\vspace{0.2cm}

\noindent \textbf{RQ8: What are the benefits of LCD?}\\
To answer RQ8, we used the Constant Comparison method to identify the benefits of LCD from the practitioners' perspective. We finally got 16 advantages of using LCD refined from 210 instances as listed in Table \ref{Benefits of LCD with their examples and counts}.

Most of the developers remarked that LCD allows for \textit{faster development}, thus rapidly brings applications to market. They also commented that LCD's \textit{ease of study and use} with \textit{lower IT costs} compared to employing developers to code the applications makes it great. Equipped with \textit{rich and ready-to-use units}, some developers reviewed that LCD is \textit{newbie friendly} as it ``\textit{helps non technical people to create software in an easy and familiar way}''. Others hold the views that LCD builds applications with \textit{improved system quality}, has \textit{strong integration and expansion capability}, requires \textit{minimal effort}, and has \textit{better customization}. A few practitioners thought that LCD is \textit{perceptually intuitive}, has \textit{superior usability} to be ``\textit{completely flexible and capable of doing almost anything}'', and always provides ``\textit{a friendly interface}'' and ``\textit{amazing UI}'' giving \textit{better user experience}. Moreover, several posts commented that LCD can \textit{deploy the app easily}, and it is quite \textit{cost-effective} with \textit{improved IT governance} compared to programming. One developer considered that LCD \textit{suits team development}.

\begin{table}[h]
\caption{Benefits of LCD with their examples and counts}
\label{Benefits of LCD with their examples and counts}
\resizebox{\columnwidth}{!}{
\begin{tabular}{|l|l|c|}
\hline
\textbf{Benefit}                                                                       & \textbf{Example}                                                                                                                                                                                                     & \textbf{Count}                       \\ \hline
Faster development                                                                     & \textit{rapidly bring apps to market}                                                                                                                                                                                & 65                                   \\ \hline
Ease of study and use                                                                  & \textit{\begin{tabular}[c]{@{}l@{}}the learning curve is very low and you can \\ start modelling very fast\end{tabular}}                                                                                     & 52                                   \\ \hline
Lower IT costs                                                                         & \textit{\begin{tabular}[c]{@{}l@{}}as the Open Edition is free, you can \\ evaluate the service or run your open-\\ source app without any financial \\ investments\end{tabular}}                                   & 18                                  \\ \hline
\begin{tabular}[c]{@{}l@{}}Rich and ready-to-use \\ units\end{tabular}                 & \textit{tons of modules and easy installation}                                                                                                                                                                      & 14                                   \\ \hline
Newbie friendly                                                                        & \textit{\begin{tabular}[c]{@{}l@{}}helps non technical people to create \\ software in an easy and familiar way\end{tabular}}                                                                                       & 10                                   \\ \hline
Improved system quality                                                                       & \textit{\begin{tabular}[c]{@{}l@{}}yields high-quality, secure, scalable and \\ maintainable applications\end{tabular}}                                                                                              & 7                    \\ \hline
\begin{tabular}[c]{@{}l@{}}Strong integration and \\ expansion capability\end{tabular} & \textit{\begin{tabular}[c]{@{}l@{}}completely integrated with Drools and \\ Drools fusion, allowing you to model and \\ execute complex business scenarios\end{tabular}}                                            & 7                                  \\ \hline
Minimal effort                                                                         & \textit{\begin{tabular}[c]{@{}l@{}}saves you tons of efforts by avoiding you \\ writing boilerplate code\end{tabular}}                                                                                               & 7                           \\ \hline
\begin{tabular}[c]{@{}l@{}}Better customization\end{tabular}                           & \textit{\begin{tabular}[c]{@{}l@{}}can be customized and scaled effortlessly \\ based on customer requirements\end{tabular}}                                                                                        & 6                         \\ \hline
Perceptual intuition                                                                   & \textit{pretty responsive out of the box}                                                                                                                                                                            & 5                                    \\ \hline
Superior usability                                                                     & \textit{\begin{tabular}[c]{@{}l@{}}It aims to be completely flexible and \\ capable of doing almost anything\end{tabular}}                                                                                          & 5                           \\ \hline
Better user experience                                                                 & \textit{\begin{tabular}[c]{@{}l@{}}provides an easy to use graphical tool \\ inside the designer that provides the ability \\ to easily transform data, create messages, \\ variables, conditions, etc\end{tabular}} & 4                             \\ \hline
Easy deployment                                                                        & \textit{\begin{tabular}[c]{@{}l@{}}makes it easier to deploy the app in mobile \\ devices (iOS / Android)\end{tabular}}                                                                                              & 4                       \\ \hline
Cost-effectiveness                                                                     & \textit{\begin{tabular}[c]{@{}l@{}}cost-effective way of developing \\ applications\end{tabular}}                                                                                                                    & 3                                   \\ \hline
Improved IT governance                                                                 & \textit{\begin{tabular}[c]{@{}l@{}}Business experts not having to rely 100\% \\ on software devs makes things different\end{tabular}}                                                                                & 2                                    \\ \hline
Improved team development                                                                  & \textit{It suits team development well}                                                                                                                                                                              & 1                                   \\ \hline
\end{tabular}
}
\end{table}

\vspace{0.2cm}

\noindent \textbf{RQ9: What are the limitations and challenges of LCD?}\\
To answer RQ9, we collected the existing shortcomings and future challenges of LCD from the selected posts and obtained 15 limitations and challenges as presented in Table \ref{Limitations and challenges of LCD with their examples and counts}.

Most practitioners reflected that although using LCD is easier and faster than coding, it still has a \textit{steep learning curve} to some degree. Some LCD platforms' \textit{high pricing} makes the cost of LCD expensive, especially if you have a large number of users. Due to restrictive customization, practitioners thought that LCD platforms \textit{lack of customization}. In addition, developers complained about \textit{slow loading and publishing} of some LCD platforms. Also, practitioners felt LCD \textit{less powerful than programming} with \textit{high complexity} to some extent. Several developers pointed out the shortcomings of LCD that \textit{complex issues still need coding}, \textit{no access of source code}, \textit{not really ease of use} as well as \textit{limitation to experienced developers} . \textit{Vendor lock-in}, \textit{difficulty of maintenance and debugging}, \textit{difficulty of integration} are also the weaknesses of LCD. Two developers stated that they had \textit{unfriendly user experience} with LCD, and two others strengthened the \textit{need of basic programming knowledge} since ``\textit{most of them do require code at some point}''.

\begin{table}[h]
\caption{Limitations and challenges of LCD with their examples and counts}
\label{Limitations and challenges of LCD with their examples and counts}
\resizebox{\columnwidth}{!}{
\begin{tabular}{|l|l|c|}
\hline
\textbf{\begin{tabular}[c]{@{}l@{}}Limitations and challenges \\ of LCD\end{tabular}}  & \textbf{Example}                                                                                                                                                                                                                                                                    & \textbf{Count} \\ \hline
High learning curve                                                                    & \textit{\begin{tabular}[c]{@{}l@{}}you need to learn a lot about how this tool \\ works to do the thing you’re trying to do\end{tabular}}                                                                                                                                          & 21                                  \\ \hline
High pricing                                                                           & \textit{\begin{tabular}[c]{@{}l@{}}These larger vendors can get expensive, \\ because they charge you for every user and you \\ have to buy packages of 50 or 100 users\end{tabular}}                                                                                              & 13                                  \\ \hline
Lack of customization                                                                  & \textit{\begin{tabular}[c]{@{}l@{}}Restrictive customisation on design and \\ layouts\end{tabular}}                                                                                                                                                                                 & 11                                  \\ \hline
Slow loading and publishing                                                            & \textit{Loading speeds can be slow}                                                                                                                                                                                                                                                 & 9                                   \\ \hline
\begin{tabular}[c]{@{}l@{}}Less powerful than \\ programming\end{tabular}              & \textit{\begin{tabular}[c]{@{}l@{}}A full-fledged programming language will \\ always have more power than a "no-\\ code/low-code" solution such as PowerApps\end{tabular}}                                                                                                        & 6              \\ \hline
High complexity                                                                        & \textit{they’re often too convoluted to use}                                                                                                                                                                                                                                        & 6                                   \\ \hline
\begin{tabular}[c]{@{}l@{}}Complex issues still need \\ coding\end{tabular}            & \textit{\begin{tabular}[c]{@{}l@{}}If you go further and having a complex issue \\ that can only be solved with invoking code or \\ creating custom activities, you really need to \\ code\end{tabular}}                                                                           & 5                                   \\ \hline
No access of source code                                                               & \textit{\begin{tabular}[c]{@{}l@{}}Therefore you cannot take the code and use it \\ elsewhere\end{tabular}}                                                                                                                                                                        & 4                                   \\ \hline
Not really ease of use                                                                 & \textit{\begin{tabular}[c]{@{}l@{}}No code is great, but not as easy as picking \\ an app that's already written\end{tabular}}                                                                                                                                                     & 4                                   \\ \hline
\begin{tabular}[c]{@{}l@{}}Limitation to experienced \\ developers\end{tabular}        & \textit{\begin{tabular}[c]{@{}l@{}}Most no-code tools are designed more like a \\ prototyping tool and also targeted for non-\\ developers which makes it very difficult for \\ someone with development background to use\end{tabular}}                                           & 4                                   \\ \hline
Vendor lock-in                                                                          & \textit{\begin{tabular}[c]{@{}l@{}}Then there's the issue of vendor lock in. If \\ you build using a nocode tool and they host \\ etc. then if they raise their prices or shut \\ down, that's going to a huge cost in \\ downtime or rebuild and possibly lost data\end{tabular}} & 3                                   \\ \hline
\begin{tabular}[c]{@{}l@{}}Difficulty of maintenance \\ and debugging\end{tabular} & \textit{\begin{tabular}[c]{@{}l@{}}An additional risk is the continued support \\ and maintenance of the low-code platform\end{tabular}}                                                                                                                                            & 3                                   \\ \hline
Difficulty of integration                                                              & \textit{\begin{tabular}[c]{@{}l@{}}it looks to be a hard problem to make the UI, \\ data store and calculations work together\end{tabular}}                                                                                                                                         & 3                                   \\ \hline
Unfriendly user experience                                                             & \textit{\begin{tabular}[c]{@{}l@{}}it has a steeper and at times user unfriendly \\ UX\end{tabular}}                                                                                                                                                                               & 2                                   \\ \hline
\begin{tabular}[c]{@{}l@{}}Need of basic programming \\ knowledge\end{tabular}         & \textit{most of them do require code at some point}                                                                                                                                                                                                                                 & 2                                   \\ \hline
\end{tabular}
}
\end{table}



\section{Discussion}
\label{sec:discussion}


\subsection{Interpretation of Results}
\noindent\textbf{\large RQ1: Terms that practitioners use to describe LCD}

The result of RQ1 (see Table \ref{Terms that practitioners used to describe LCD}) shows that most practitioners tend to use the term \textit{low-code} to describe what is LCD. \textit{Low-code} here means creating software with radically small amounts of code, or even without hand-coding. There are some other practitioners who use terms, such as \textit{drag and drop}, \textit{visual programming}, and \textit{graphical user interface} when referring to LCD. This indicates that LCD probably provides users with a GUI to point and click, the frontend layout, the backend logic, or even the connection to the third party APIs. Moreover, some others define LCD as the equipment of \textit{pre-designed templates}, which is a higher-level implementation unit than code and is easier to learn and use. On account of LCD to save code effort, WYSIWYG to build everything visually, and plenty of well-built templates to develop applications with, we can explain why practitioners also hold the point of view that LCD is \textit{non-professional programmers friendly}. 

\noindent\textbf{\large RQ2: Platforms that are used in LCD} 

For RQ2, we listed the platforms that were mentioned at least 10 times in the 301 posts in Table \ref{Platforms used in LCD, their companies, whether open-source or not and counts}. We found that some of these platforms are \textit{commercial}, while others are \textit{open-source}. On the one hand, given that developers are concerned about the problems of \textit{vendor lock-in} and \textit{no access of source code}, they would better choose open-source platforms to own the code that they have control and do not need to worry about that vendors may raise their prices or shut down the platforms one day. On the other hand, commercial software may offer more advanced and convenient functions, while open source software may require less financial support and is suitable when there is a tight budget. 
We also categorized all the LCD platforms appeared in the posts according to the parts of applications developed by LCD as shown in Table \ref{Platforms classified by the parts of applications developed by LCD}. The result shows that the same platform may support the development of various parts of an application, which potentially extends LCD platforms to an integration platform, for example, from \textit{frontend}, \textit{workflow} to \textit{integration}. 

\noindent\textbf{\large RQ3: Programming languages that are used in LCD} 

The result of RQ3, to some degree, depends on the supporting technologies behind LCD, which are largely specific to the LCD platform chosen. Figure \ref{fig:Programming language used in LCD} shows that \textit{Java} and \textit{JavaScript} are both popular programming languages used for LCD, which is reasonable in that they are also popular languages in software development. 

\noindent\textbf{\large RQ4: Implementation units which LCD is based on} 

LCD is called ``low-code'' since low-code development is based on higher level implementation units than code. These out-of-the-box units equipped in LCD platforms play an important role in making platforms easy to use and speeding up the development. The major implementation units an LCD platform uses usually depend on the parts of applications developed by LCD. Figure \ref{fig:Implementation Unit in LCD} shows that most of the LCD platforms use \textit{APIs}. After further checking the data extracted, we found that the implementation units \textit{APIs} and \textit{services} are usually used to develop the \textit{backend} of applications, while the implementation units \textit{components} and \textit{widgets} are used to develop the \textit{frontend} of applications. The implementation units \textit{templates} and \textit{frameworks} are often employed for developing both the \textit{frontend} and \textit{backend} of applications. The implementation units are perceptually more intuitive than code, thus users are not aware of the complexities with those well-encapsulated units. We believe that these basic implementation units of LCD can definitely make development more agile and the LCD platforms easier to use.

\noindent\textbf{\large RQ5: Supporting technologies behind LCD} 

For those supporting technologies behind LCD, \textit{React Native} is an open-source mobile application framework, and it serves as a concrete implementation unit in several LCD platforms. \textit{Node.js} is a free, open-sourced, cross-platform JavaScript run-time environment that allows developers to write command line tools and server-side scripts outside of a browser, and the post implies that it may support drag and drop interfaces of LCD platforms. \textit{Apache Spark} is a unified analytics engine for large-scale data processing, and it makes LCD capable of data processing and analysis. For example, Mapping data flows in Azure Data Factory run Apache Spark underneath to allow data engineers to develop data transformation logic without writing code. \textit{Server-side scripting} is a technique used in web development which involves employing scripts on a web server and produces a response customized for each user's request to the website, and it is a kind of technology used behind drag and drop website builder allowing users to create a website with zero code. \textit{SOAP} (i.e., Simple Object Access Protocol) is a messaging protocol specification for exchanging structured information in the implementation of web services. The LCD platform Bonita offers a SOAP connector for consuming SOAP web services. \textit{JSONata} is a declarative open-source query and transformation language for JSON data, and it powers the magic behind nearly all LCD platforms to translate data schemas between platforms. The supporting technologies behind LCD are diverse without the dominant, which shows that LCD is not constrained to specific technologies.

\noindent\textbf{\large RQ6: Types of applications developed by LCD} 

Many posts state the types of applications which developers created or were creating using LCD. From the result, the demand for developing \textit{mobile applications} with low code is the highest among the types of applications, which is reasonable since mobile applications normally have a short delivery time and their development can be sped up by employing LCD. Different LCD platforms support the development of different types of applications. For instance, Bubble introduces a way to build web applications without code, Storyboard is used to create interfaces for iOS apps with zero code, and Syndesis is an open source integration platform that can connect to any services and provides a rich set of connectors out of the box. Therefore, before development, it is necessary to choose appropriate LCD platforms for the projects according to the needs of the type of applications developed.

\noindent\textbf{\large RQ7: Domains that use LCD} 

The result of the domains that use LCD is presented in Figure \ref{fig:Domain}. Applications belonging to \textit{BPM}, \textit{CRM}, \textit{CMS}, and \textit{RPA} are often software systems for workflow management or business process automation. Furthermore, Table \ref{Platforms classified by the parts of applications developed by LCD} shows that 44 of the 137 platforms support workflow development, and many platforms have mature implementation units to build automation. Besides, most developers were concerned about whether they can build \textit{E-commerce} applications through LCD, which are also process intensive.

\noindent\textbf{\large RQ8 and RQ9: Benefits brought by LCD and limitations and challenges of LCD} 

LCD does allow users to build applications faster with minimal effort. It makes development more agile as LCD platforms are equipped with rich and ready-to-use implementation units, but this also leads to the slowness of loading and publishing. Besides, the downsides of LCD include no access of source code and vendor lock-in. However, we found that many benefits contradict its limitations and challenges from the practitioners' perspective: 
\textbf{(1)} Many developers think that LCD platforms are \textit{easy to study and use} compared to programming languages, however, some developers also argue that they are \textit{not really ease of use} with \textit{high learning curve}. LCD is indeed easier than programming, but the use of LCD platforms also has a certain learning cost, especially some LCD platforms provide complex functions, which take time to learn.
\textbf{(2)} Some practitioners also consider \textit{lower IT costs} than traditional development as a benefit brought by LCD, but others hold the opposite view that using LCD is \textit{expensive}. On the one hand, LCD reduces the time required for development and allows non-professional to implement their ideas without having to hire developers; on the other hand, some commercial LCD platforms also require a high price to provide a complete service, and some of the platforms charge for every user, which means that they can get very expensive as you scale your team. 
\textbf{(3)} LCD is \textit{newbie-friendly} in some developers' point of view while \textit{limiting experienced developers} to others, as some LCD platforms are designed more like a prototyping tool and also target for non-developers which makes it very difficult for \textit{experienced developers} to use.
\textbf{(4)} In some practitioners' perspective, LCD yields \textit{high-quality, secure, scalable, and maintainable applications} based on out-of-the-box implementation units; in contrast, a few believes that LCD solutions are \textit{hard to modify, maintain, and debug}. The potential reason could be that it is difficult to verify whether the implementation units provided by LCD platforms have defects due to no access of code, and different LCD platforms may bring different experiences to developers. 
\textbf{(5)} Some LCD platforms are considered to have \textit{perfect customization} while others complain about the \textit{lack of customization}. An LCD platform with limited flexibility in functionality and design may result in spending more time to add custom code unless it gives the users enough feature set to customize.
\textbf{(6)} LCD seems \textit{intuitive} if the users only build applications with drag and drop operations. Once they find it limited and inflexible and have to add custom code at some point, LCD may bring \textit{complexities}, compromises, and frustrations. Furthermore, if \textit{complex functions still need coding} to achieve, it means that LCD is \textit{less powerful than programming}.

\subsection{Implications}

\noindent\textbf{Definition of LCD:} Although the term ``low-code'' was first coined in 2014 \cite{richardson2014new}, to date, there is no clear definition of ``low-code'' in either academia or industry. Practitioners use a variety of terms to describe low-code related practices, and we suggest that researchers refine and summarize these descriptions to reach a level of consensus in order to reduce misunderstandings and ambiguities. 
We also hope that practitioners will gradually deepen their understanding of LCD along with the practice in development, and make the concept and scope of LCD clear.


\noindent\textbf{Choice of platforms in LCD:}
When developers seek help on SO and Reddit, some tend to ask questions like ``\textit{Suggest a BPM like tool to automate unattended operations with minimal human involvement}'' or ``\textit{Can you use Microsoft Power Automate to develop a company wide workflow solution?}'', suggesting these developers are unaware of which LCD platform is appropriate for their needs. Once a decision has been made to build an application in a low-code way, the developers should choose a suitable platform, taking into account the characteristics of the platforms and their own needs. From the perspective of platform features, developers need to consider whether they want to have access to the source code and deploy the code on their own servers. They should also take the budgets into account to choose a platform with reasonable pricing. Furthermore, developers may choose an LCD platform that supports the programming language they are good at. Finally, it is of vital importance to know whether the platform they choose provides enough implementation units to support the parts of applications, e.g., frontend UI, backend logic, and data store, to be developed using LCD. From the perspective of user requirements, developers should choose the right platform based on the type and domain of the application they want to develop. 

\noindent\textbf{Adoption of LCD in projects:}
LCD proposes a new programming paradigm, and from the interpretation of the results of RQ8 and RQ9, we can see that it cannot replace the traditional development approaches. Developers should consider the estimated time of development cycle, learning cost, the budget, the extent of customization provided, and some other aspects carefully, to decide whether they are supposed to use the LCD approach with an appropriate LCD platform chosen or adopt a traditional development approach with hand-coding and maximal human involvement. In short, LCD is like a double-edged sword. If the features of LCD are well used to meet the requirements of development, they will certainly accelerate the development process. Otherwise, they will increase the degree of difficulty and complexity of development, making non-professional developers feel frustrated and professional developers feel constrained.

\section{Threats to Validity}
\label{sec:threats}
We discuss the potential threats to the validity of our results below by following the guideline in \cite{wohlin2012book}. Internal validity is not discussed, since we did not investigate any causal relationships.

\textbf{\large Construct validity} denotes whether the theoretical and conceptual
constructs are correctly measured and interpreted. In our study, there are three threats to this validity: (1) One threat stems from the selection of data sources. Since many LCD platforms have their own forums, it it possible that some LCD platforms may not have much relevant discussion in SO and Reddit, which might bring bias in dataset. (2) Another threat is related to the search terms we used to collect posts from SO. Our selected search terms may not provide a comprehensive coverage of all posts related to LCD. To mitigate this threat, we have reviewed the relevant literature and searched the synonyms of LCD on the Internet. (3) The last threat comes from manual extraction and analysis of data. To mitigate the impact of this threat, we randomly selected posts from the dataset and did a pilot data extraction and analysis. In the formal data extraction and analysis process, any uncertainty was discussed by the first, second, and fifth authors until an agreement was reached to eliminate personal bias.
 
\textbf{\large External validity} concerns the extent to which the results of a study can be generalized to and across other situations, people, settings, and measures. A relevant threat concerns the selection of online developer communities. To reduce this threat, in our study, we collected data from two data sources Stack Overflow and Reddit, and both are popular online development communities, which partially mitigates this threat.

\textbf{\large Reliability:} refers to the replicability of a study for arriving at same or similar results. To alleviate this threat, we defined a research protocol with detailed procedure which can be used to reproduce our work and it was discussed and confirmed by all the authors; our research process is explicitly shown in the Methodology section (Section \ref{sec:methodology}), and the dataset and coding results from the study have been made available online \cite{replpack}. Furthermore, before the formal data extraction, we conducted a pilot data extraction between the first, second, and fifth authors until reaching a consistent understanding about the extraction results. With these measures, we are confident that the study results are relatively reliable.

\section{Conclusions} 
\label{sec:conclusion}
LCD is not a new concept in software development, but recently is booming and gets much attention in industry that can help both developers and users quickly deliver applications by writing a few code. We conducted an empirical study to obtain the characteristics as well as the challenges towards LCD from a practitioners’ perspective. We used LCD related terms to search and collect data from SO and used LCD related subreddits from Reddit to collect data for data extraction and analysis with 301 posts in total. The main findings of this study are the following:

\begin{itemize}
    \item Although there is no clear definition of LCD, practitioners tend to use \textit{low-code} and \textit{drag and drop} to describe LCD according to their understanding, showing that LCD may provide a graphical user interface for users to drag and drop with little or even no code.
    \item The equipment of out-of-the-box units in LCD platforms makes them easy to learn and use as well as speeds up the development.
    \item Different LCD platforms support the development of different types (e.g., mobile and web applications) of applications and different parts (e.g., frontend, workflow, and integration) of applications.
    \item LCD is particularly favored in the domains that have the need for automated processes and workflows.
    \item While LCD platforms can speed up software development with minimal human involvement, they also suffer from no access of source code and vendor lock-in for commercial LCD platforms. Moreover, practitioners have conflicting views on the advantages and disadvantages of LCD, implying that certain features of LCD are beneficial to development if used appropriately, otherwise may become limitations or challenges in LCD. Therefore, developers should consider whether LCD is appropriate for their projects.
\end{itemize}

In the next step, we plan to extend this work on studying LCD in a larger dataset from multiple developer communities and methods (e.g., using questionnaire and interview). We also intend to take a deeper look at various aspects of LCD, such as in which condition, a benefit becomes a limitation in LCD, and how to prevent that.

\begin{acks}
This work has been partially supported by the National Key R\&D Program of China with Grant No. 2018YFB1402800.
\end{acks}

\balance
\bibliographystyle{ACM-Reference-Format}
\bibliography{references}

\end{document}